\documentclass[nofootinbib,twocolumn,showpacs,preprintnumbers,pre,aps,superscriptaddress]{revtex4-2}

\usepackage[dvipdfmx]{graphicx}% Include figure files
\usepackage{bm}% bold math
\usepackage{amsmath}
\usepackage{amssymb}
\usepackage{newtxtext}
\usepackage{newtxmath}
\usepackage{color}
\usepackage{hyperref}
\usepackage{tikz}
\usepackage{float} 
\allowdisplaybreaks[1]

\begin{document}
%%%%%%%%%%%%%%%%%%%%%%%%%%%%%%%%%%%%%%%
\title{Domain growth in lipid membranes and the budding instability}
%%%%%%%%%%%%%%%%%%%%%%%%%%%%%%%%%%%%%%%

\author{Jean Wolff}
\email{jean.wolff@ics-cnrs.unistra.fr}

\author{Fabrice Thalmann}\email{fabrice.thalmann@ics-cnrs.unistra.fr}
\affiliation{Institut Charles Sadron, CNRS UPR22 \& Universit\'{e} de Strasbourg, Strasbourg 67000, France}

\author{Carlos M. Marques}\email{carlos.marques@ens-lyon.fr}
\affiliation{University of Lyon, ENS-Lyon, CNRS UMR 5182, Chemistry Laboratory, Lyon 69342, France}

\author{David Andelman}\email{andelman@tauex.tau.ac.il}
\affiliation{Center for Physics and Chemistry of Living Systems, 
Tel Aviv University, Ramat Aviv 6997801, Tel Aviv, Israel}

\author{Haim Diamant}\email{hdiamant@tau.ac.il}
\affiliation{School of Chemistry \& Center for Physics and Chemistry of Living Systems, 
Tel Aviv University, Ramat Aviv 6997801, Tel Aviv, Israel}

\author{Shigeyuki Komura}\email{komura@wiucas.ac.cn}
\affiliation{Zhejiang Key Laboratory of Soft Matter Biomedical Materials, Wenzhou Institute, 
University of Chinese Academy of Sciences, Wenzhou, Zhejiang 325001, China}

\date{2026/09/26}
% Version 22 submitted to BJ

%%%%%%%%%%%%%%%%%%%%%%%%%%%%%%%%%%%%%%%
\begin{abstract}
We investigate the growth of a circular domain in a phase-separating lipid membrane and its relation to the 
budding instability. 
We describe the in-plane dynamics with a conserved, time-dependent Ginzburg-Landau model under radial symmetry. 
Numerical solutions show a crossover of the domain size $D$ as a function of time $t$ from $D\sim t^{1/3}$ at early times to $D\sim t^{1/2}$ at later times, 
followed by saturation due to finite system size. 
The time-dependent line tension $\sigma(t)$, evaluated from the evolving concentration profile, rises rapidly before 
approaching the planar-interface value. 
This behavior distinguishes an early line-tension-dominated regime from a later domain-size-dominated regime. 
To estimate the budding instability, we combine $D(t)$ and $\sigma(t)$ with a quasistatic criterion for the loss of stability 
of a partially budded domain.
The budding-instability time decreases with increasing membrane spontaneous curvature and increases near the critical point. 
Thus, the evolution of both domain size and line tension is essential for determining the time required for the budding instability.
\end{abstract}
%%%%%%%%%%%%%%%%%%%%%%%%%%%%%%%%%%%%%%%%%%%%%%%%%%%%%%%%%%%
\maketitle
%\tableofcontents

%%%%%%%%%%%%%
\section{Introduction}
%%%%%%%%%%%%%
\label{sec:introduction}

%%%%%%%%%%%%%%%%%%%%%%%%%%%%%%%%%%%%%%
\begin{figure}[tbh]
\centering
\includegraphics[width=0.75\linewidth]{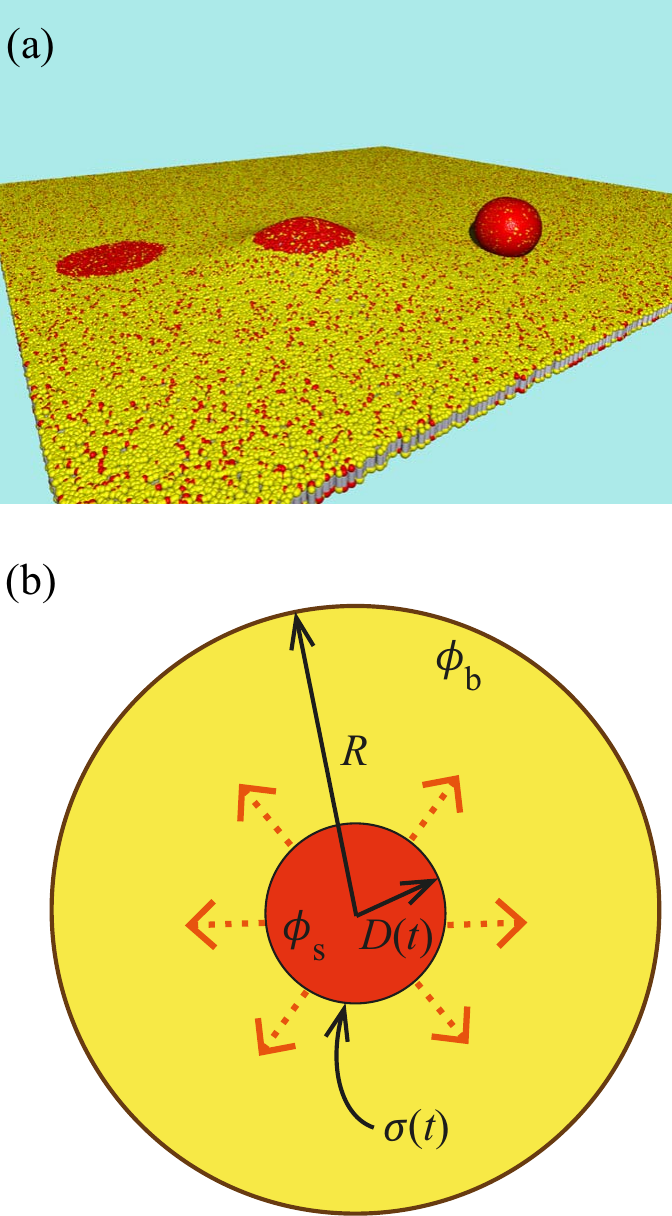}
\caption{
\textsf{
(a) Schematic illustration of a planar or dimpled domain (red) that buds from an otherwise planar two-dimensional (2D) membrane (predominantly yellow). 
(b) Schematic representation of the model. 
The inner domain with the stable-phase composition $\phi_{\rm s}$ (red) grows into the surrounding background with initial composition $\phi_{\rm b}$ (yellow). 
The time-dependent radius of the growing inner domain is denoted by $D(t)$, while $R$ is the outer radius of the membrane. 
The boundary of the growing inner domain, located at $r=D(t)$, is characterized by the line tension $\sigma(t)$.
}}
\label{fig1}
\end{figure}
%%%%%%%%%%%%%%%%%%%%%%%%%%%%%%%%%%%%%%

Biological membranes are fluid bilayers composed mainly of phospholipids and cholesterol~\cite{deMeyer2009}. 
They consist of two apposed leaflets~\cite{1925_Gorter_Grendel,1972_Singer_Nicolson} and have a typical thickness of about 
$5$\,nm~\cite{2000_Petrache_Brown}. 
In each leaflet, hydrophilic headgroups face the aqueous environment, whereas hydrophobic 
tails form the membrane interior~\cite{2006_Dimova_Lipowsky}. 
In aqueous solution, these amphiphilic molecules self-assemble into bilayers and can form \textit{in vitro} vesicles~\cite{2010_Loudet_Dufourc}.
Changes in temperature or lipid composition can induce lateral liquid-liquid phase separation and the formation of lipid 
domains~\cite{1996_Widom,2007_Yanagisawa,2006_Saeki,2013_Orlandini, 2011_Wolff}. 
Because the membrane remains fluid, these domains evolve in shape and size. 
Approximately circular domains are commonly observed during in-plane growth, reflecting the tendency to reduce 
the length of the boundary between coexisting phases.

Under suitable conditions, membrane domains may deform out of the membrane plane and undergo budding~\cite{2010_Hurley,2024_Wu}. 
This process occurs in both artificial vesicles and biological membranes. 
In cells, budding is involved in transport, signaling, and the exchange of material across membranes, 
while in engineered systems it can help control vesicle morphology and liposome size~\cite{2011_Bae}. 
Understanding the onset of budding, therefore, requires linking lateral phase-separation dynamics to membrane-deformation energetics.

Lipowsky~\cite{1992_Lipowsky} proposed an equilibrium mechanism for the budding of a planar or weakly curved circular domain~\cite{2005_McMahon}. 
Above a critical size, a two-dimensional (2D) domain may form either a complete bud or a dimple, 
i.e., a partially budded state that remains connected to the surrounding membrane [see Fig.~\ref{fig1}(a)]. 
The instability is governed by the competition between the domain boundary line energy and the membrane's bending energy.
The corresponding criterion depends on the domain radius $D$, line tension $\sigma$, bending rigidity, and spontaneous curvature 
[see later Eq.~\eqref{eq15}].
Related equilibrium studies have shown that line tension, membrane tension,
and spontaneous curvature can control transitions between different domain morphologies~\cite{2005_Harden}.

The equilibrium theory, such as in Ref.~\cite{1992_Lipowsky}, neglects membrane surface tension and treats $\sigma$ as a prescribed constant. 
During lateral phase separation, however, both the domain radius and the interfacial structure evolve. 
The radius $D(t)$ increases through lipid transport, while the line tension $\sigma(t)$ changes with the concentration profile 
and domain-boundary curvature. 
The onset of budding therefore depends on the joint evolution of $D(t)$ and $\sigma(t)$ rather than on fixed equilibrium values.

In our previous work, we studied equilibrium budding in symmetric and asymmetric bilayers~\cite{2015_Wolff,2016_Wolff}.
Here, extending these studies and related work on dynamical membrane deformation~\cite{2004_Sens,2007_Hamada,2009_Franke}, 
we investigate how the nonequilibrium growth of a circular domain drives the system toward a budding instability.
We describe the in-plane phase-separation dynamics by ``Model B" for a conserved order parameter and determine 
the time-dependent domain radius $D(t)$ and line tension $\sigma(t)$ directly from the evolving concentration profile.
These two dynamical quantities are then incorporated into the equilibrium budding criterion within a quasistatic approximation, 
allowing us to estimate when a growing domain becomes unstable toward further 
budding~\cite{2008_Sakuma,2008_Yanagisawa,2009_Arroyo,2010_Camley,2015_Barrett}.

Our results reveal two distinct stages in the evolution toward budding.
The domain growth exhibits a crossover between different growth regimes, while the line tension increases 
rapidly at early times and subsequently approaches its planar-interface value.
As a result, the approach to the budding threshold is governed primarily by the evolution of the line tension 
at early times, but increasingly by domain growth at later times.
The budding instability is therefore determined by the coupled evolution of $D(t)$ and $\sigma(t)$ rather than 
by a fixed critical radius or a constant line tension alone.
Within the quasistatic description, the resulting budding-instability time decreases with increasing spontaneous 
curvature and increases as the system approaches the critical point.

The paper is organized as follows. 
In Sec.~\ref{sec:model}, we formulate a Model B description of the radial growth of a circular domain in a binary lipid membrane~\cite{1995_Chaikin}. 
In Sec.~\ref{sec:Dy}, we present the evolution of the concentration profile, domain radius, and line tension, 
including the crossover between the two growth laws. 
In Sec.~\ref{sec:lipo}, we combine $D(t)$ and $\sigma(t)$ with the budding criterion of Ref.~\cite{1992_Lipowsky} and determine 
the budding-instability time as a function of spontaneous curvature and distance from the critical point. 

%%%%%%%%%%%%%%%%%%%%%%%%
\section{Domain Growth in a Membrane}
%%%%%%%%%%%%%%%%%%%%%%%%
\label{sec:model}

%%%%%%%%%%%%%%%%%%%%%%%%%%%%%%%%%%%%
\subsection{Phase-separation dynamics in a binary membrane}
\label{sec:binary}

Consider a 2D planar fluid membrane composed of two species, A and B, as shown in Fig.~\ref{fig1}(a).
The local 2D number densities of the two lipid species are denoted by 
$n_{\rm A}(\mathbf{r},t)$ and $n_{\rm B}(\mathbf{r},t)$, which depend on 2D position 
$\mathbf{r}$ and time $t$.
Assuming equal molecular areas and local incompressibility, their total number density 
$n_{\mathrm{tot}}=n_{\rm A}+n_{\rm B}$ is uniform in space and constant in time. 
We introduce the dimensionless densities
$\phi_{\rm A}(\mathbf{r},t)=2n_{\rm A}(\mathbf{r},t)/n_{\mathrm{tot}}$ and 
$\phi_{\rm B}(\mathbf{r},t)=2n_{\rm B}(\mathbf{r},t)/n_{\mathrm{tot}}$,
which satisfy $\phi_{\rm A}+\phi_{\rm B}=2$. These densities have means $\bar\phi_{\rm A}$ and $\bar\phi_{\rm B}$.
The conserved order parameter describes the composition
$\phi(\mathbf{r},t)=\phi_{\rm A}(\mathbf{r},t)-\phi_{\rm B}(\mathbf{r},t)$, 
where $ \bar{\phi}=\bar{\phi}_{\rm A} - \bar{\phi}_{\rm B}$ is assumed to be zero in this work. 

\begin{figure}
\centering
\includegraphics[width=0.8\linewidth]{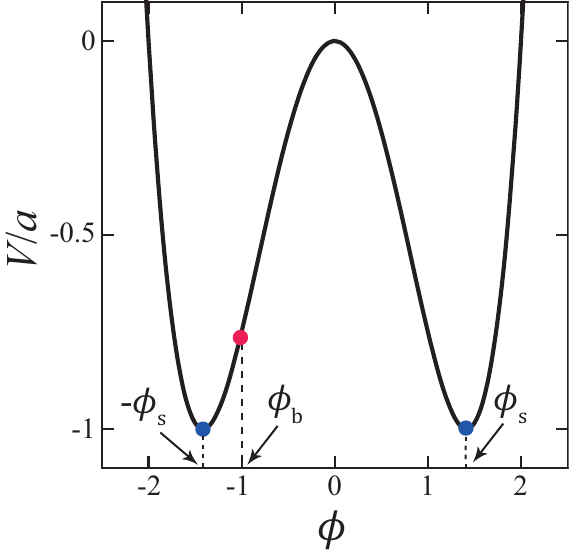}
\caption{
\textsf{
The dimensionless free-energy density $V(\phi)/a$ is plotted as a function of the 
composition $\phi$ [see Eq.~\eqref{eq2}]. 
Here, the parameter value is $a/b=2$. 
The blue dots indicate the two stable phases, $\phi=\pm\phi_{\rm s}=\pm\sqrt{a/b}=\pm\sqrt{2}$, 
while the red dot marks the chosen initial background composition, $\phi_{\rm b}=-1$.
}}
\label{fig2}
\end{figure}

To model the phase separation between the two components, we use the standard Ginzburg-Landau 
free-energy functional for liquid-liquid phase separation,
\begin{equation}
F[\phi(\mathbf{r},t)]
=\int {\rm d}^2 r \left[ \frac{c}{2}(\nabla\phi)^2+V(\phi)\right],
\label{eq1}
\end{equation}
where $c$ is the gradient-energy coefficient associated with the energetic cost of the domain 
boundary, and $V(\phi)$ is the local free-energy density,
\begin{equation}
V(\phi)= -\frac{a}{2}\phi^2+\frac{b}{4}\phi^4.
\label{eq2}
\end{equation}
Here, $a \sim T_{\rm c}-T$ measures the distance from the critical temperature $T_{\rm c}$, 
and $b>0$ is a positive constant. 
For $a<0$ ($T>T_{\rm c}$), the free-energy density has a single stable minimum, whereas for 
$a>0$ ($T<T_{\rm c}$), it has two degenerate minima separated by an unstable region. 
The dimensionless free-energy density, $V(\phi)/a$, is shown in Fig.~\ref{fig2} 
for $a/b=2$ ($T<T_{\rm c}$).
It indicates the two minima, $\phi=\pm\phi_{\rm s}=\pm \sqrt{a/b}$, together with the 
chosen initial background composition $\phi_{\rm b}$. The equilibrium values of $\phi$ in the two phases are determined by 
the two minima of $V-\mu_{\rm eq}\phi$, where $\mu_{\rm eq}$ is the equilibrium chemical potential associated with $\phi$. 
At the critical point we have $T=T_{\rm c}$ and $\mu_{\rm eq}=0$. For simplicity, we assume that the system is quenched 
by decreasing $T$ through the critical point on the $T$-$\mu_{\rm eq}$ plane. In this case, $\pm\phi_{\rm s}$ are the stable, 
equilibrium compositions of the two phases.

To discuss the dynamics of phase separation, we introduce the flux 
$\mathbf{j} = -L\nabla (\delta F/\delta\phi)$, where $L$ is a transport coefficient. 
Because $\phi$ is a conserved order parameter, it obeys the continuity equation
\begin{equation}
\frac{\partial\phi}{\partial t} + \nabla\cdot\mathbf{j}=0.
\label{eq4}
\end{equation}
Hence, we obtain the Model B dynamical equation~\cite{1995_Chaikin,1977_Hohenberg},
\begin{equation}
\frac{\partial\phi}{\partial t} = L\nabla^2 \left(
-c\nabla^2\phi-a\phi+b\phi^3 \right).
\label{eq5}
\end{equation}

Following a temperature quench, we consider a large planar lipid bilayer with an initially uniform 
background composition $\phi_{\rm b}$, as shown in Fig.~\ref{fig1}(b).
A circular nucleus whose core has the stable-phase composition $\phi_{\rm s}$ 
is placed at the center ($r=0$) and smoothly connected to the surrounding background. 
The nucleus is subsequently maintained by fixing the order parameter at the origin to $\phi(0,t)=\phi_{\rm s}$, 
as specified later.
We assume radial symmetry, so that the domain remains circular and its radius evolves as $D(t)$. 
Accordingly, the initial order-parameter profile satisfies $\phi(r,0)\approx \phi_{\rm s}$ close to $r=0$ and 
$\phi(r,0)\approx \phi_{\rm b}$ far from the nucleus, with a smooth transition across the interface. 
We model the surrounding membrane as a circular region of radius $R$. 
For the numerical calculations, we choose $a/b=2$, so that $\phi_{\rm s}=\sqrt{2}$, and set the initial background 
composition to $\phi_{\rm b}=-1$ and $R/\xi=100$, where $\xi=\sqrt{c/a}$ is the correlation length, also equal to the interfacial width.

%%%%%%%%%%%%%%%%%%%%%%%%%%%%%%%%%%%%%%
%%%%%%%%%%%%%%%%%%%%%%%%%%%%%%%%%%%%%%

In Appendix~\ref{he}, we show that membrane hydrodynamics does not contribute to the isotropic growth of a circular 
domain in the present model and can therefore be neglected. This is a known feature of phase-separation kinetics 
with hydrodynamics (Model H) for isotropic growth \cite{Onuki2002}. 
Equation~\eqref{eq5} thus provides the governing equation for the domain-growth dynamics considered here.

%%%%%%%%%%%%%%%%%%%%%%%%%%%%
\subsection{Radial growth of a circular domain}
\label{sec:circular}

Under radial symmetry, the domain growth depends only on the radial coordinate $r$. 
In polar coordinates, Eq.~\eqref{eq5} for $\phi(r,t)$ reduces to
\begin{eqnarray}
\frac{\partial\phi}{\partial t}&= & L \bigg[ -c \frac{\partial^4\phi}{\partial r^4}-
\frac{2c}{r}\frac{\partial^3\phi}{\partial r^3}+\left(\frac{c}{r^2}-a+3 b\phi^2\right)\frac{\partial^2\phi}{\partial r^2}
\nonumber\\
&+& \left(-\frac{c}{r^3}-\frac{a}{r}+\frac{3 b\phi^2}{r}\right) \frac{\partial\phi}{\partial r}
+6 b \phi\left(\frac{\partial\phi}{\partial r}\right)^2 \bigg].
\label{eq6}
\end{eqnarray}
In order to make the above equation dimensionless, we introduce the 
characteristic time $\tau=\xi^4/(Lc) =c/(La^2)$. 
In the numerical simulation, we use the dimensionless variables $r/\xi$ and $t/\tau$.
After this rescaling, Eq.~\eqref{eq6} depends only on $\phi_{\rm s}=\sqrt{a/b}$.

The radial Model B equation is fourth order in space and is solved on the finite interval
$0\leq r\leq R$. 
The radial chemical potential $\mu=\delta F/\delta\phi$ is given by 
\begin{equation}
\mu(r,t) = -c\left( \frac{\partial^2\phi}{\partial r^2} +\frac{1}{r}\frac{\partial\phi}{\partial r} \right)
-a\phi+b\phi^3, 
\label{CP}
\end{equation}
and the radial flux is given by $j_r(r,t)=-L \partial\mu/\partial r$.
At the origin, $r=0$, we impose
\begin{equation}
\phi(0,t)=\phi_{\rm s},
\qquad
j_r(0,t)=0.
\label{BC1}
\end{equation}
The first condition maintains the pre-existing stable nucleus by pinning the order parameter at its 
center, while the second expresses the absence of a flux through the origin. 

At the outer boundary $r=R$, we impose
\begin{equation}
\left. \frac{\partial\phi}{\partial r}\right|_{r=R}=0,
\qquad
j_r(R,t)=0.
\label{eq8} 
\end{equation}
The first condition represents a neutral outer boundary without preferential wetting, whereas the 
second prevents material exchange with the surroundings. 
The no-flux condition at $r=R$ ensures conservation of the total order parameter within the finite 
circular membrane. 

Together with the four boundary conditions given above, the time evolution is specified by the initial profile 
chosen as 
\begin{equation}
\phi(r,0) = \phi_{\rm s} +\frac{\phi_{\rm b}-\phi_{\rm s}}{2} 
\left[1+\tanh\left(\frac{r-\epsilon}{\xi} \right)\right],
\label{initial} 
\end{equation}
such that, $\phi(0,0)\approx \phi_s$ by choosing $\epsilon/\xi=3$, 
and $\phi(R,0)\approx \phi_b$ noting that $\epsilon\ll R$. 

%%%%%%%%%%%%%%%%%%%%%%%%%%%%%%%%%%%%%%
\begin{figure*}[tbh]
\centering
\includegraphics[width=0.9\linewidth]{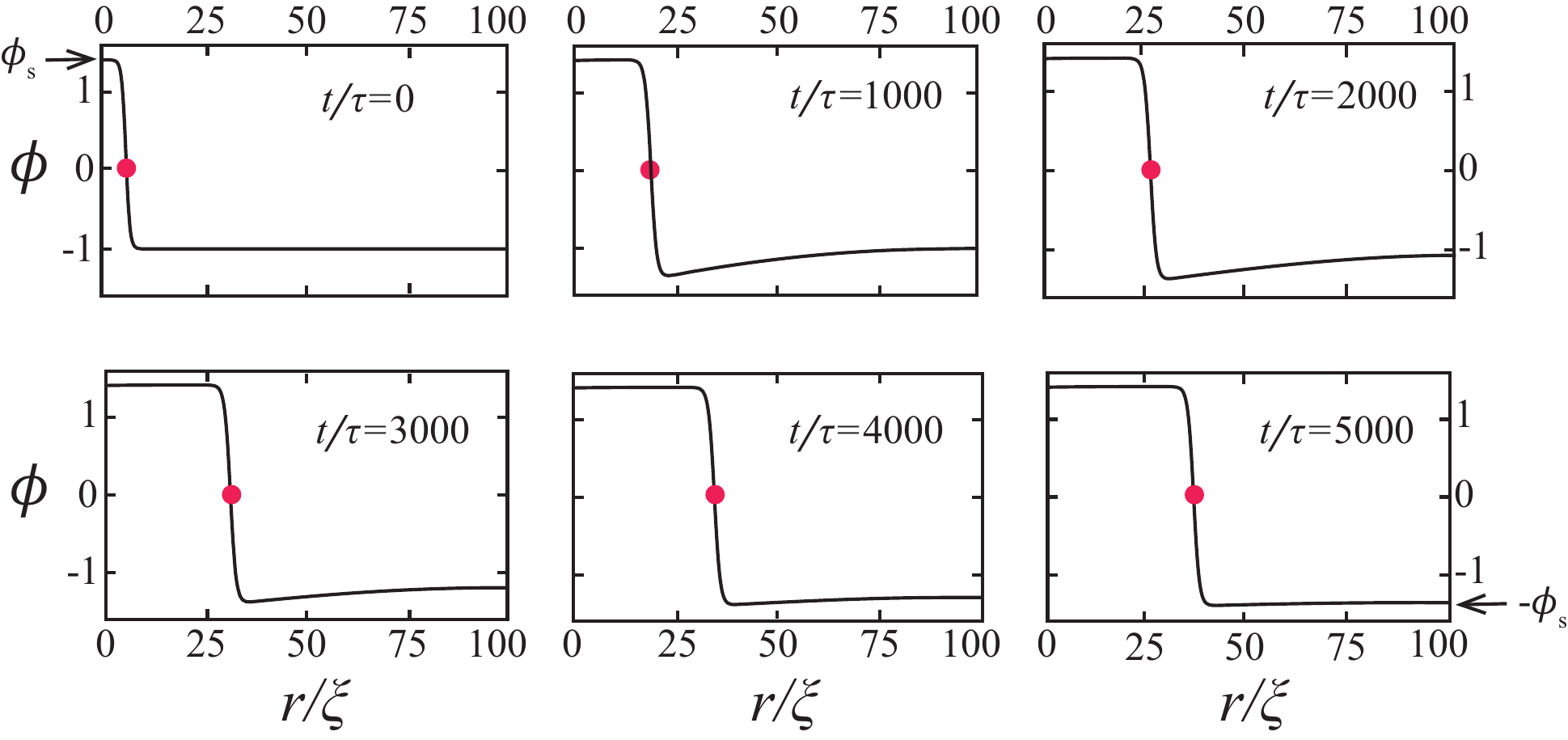} 
\caption{\textsf{
Time evolution of the composition profile $\phi(r,t)$ as a function of $r/\xi$. 
The chosen parameters are $a/b=\phi_{\rm s}^2=2$, $\phi_{\rm b}=-1$, and $R/\xi=100$.
The boundary conditions are 
$\phi(0,t)=\phi_{\rm s}$ and $j_r(0,t)=0$ at $r=0$ [see Eq.~\eqref{BC1}],
and 
$\partial_r \phi(R,t)=0$ and $j_r(R,t)=0$ at $r=R$ [see Eq.~\eqref{eq8}].
The inflection point of the profile, indicated by the red dot, defines the domain radius $D(t)$, 
which increases with time. 
}}
\label{fig3}
\end{figure*}
%%%%%%%%%%%%%%%%%%%%%%%%%%%%%%%%%%%%%%

%%%%%%%%%%%%%%%%%%%%%
\section{Domain Growth Dynamics}
%%%%%%%%%%%%%%%%%%%%%
\label{sec:Dy}

%%%%%%%%%%%%%%%%%%%%%%%%%%%%%%%%
\subsection{Time evolution of the concentration profile}
\label{sec:profile}

Figure~\ref{fig3} shows the time evolution of the concentration profile for 
$0 \leq t/\tau \leq 5{,}000$.
As seen in the figure, the profiles are nonmonotonic and exhibit an undershoot below $\phi_{\rm b}$. 
This behavior arises from the conserved dynamics together with the zero-flux boundary condition at 
$r=R$ in Eq.~\eqref{eq8}. 
Physically, material must be redistributed from the surrounding outer region to support the growth of the inner domain.
We also find that the concentration at the membrane edge, $\phi(R,t)$, 
evolves from its initial value $\phi_{\rm b}=-1$ (as we can see in Eq.~\eqref{initial}   considering $R \gg \epsilon$) and 
asymptotically approaches the equilibrium outer-phase value, $-\phi_{\rm s}=-\sqrt{2}$, as $t\to\infty$ (as we can see in Eq.~\eqref{CP} 
considering that the chemical potential is constant at $\mu_{\rm eq}=0$, and all the derivatives of $\phi$ are equal to zero).

The time-dependent radius of the inner domain, $D(t)$, is defined as the position of the inflection point of the concentration profile, 
where $\partial_r^2\phi=0$; this position is indicated by the red dot in Fig.~\ref{fig3}.
As shown in the figure, $D(t)$ increases with time and approaches a plateau value, $D_{\rm sat}$, for $t/\tau\gg1$.
In an infinite system, the domain radius would continue to grow at late times and could become macroscopically large.
In the finite system considered here, however, the outer radius $R$ is fixed, so that the long-time domain 
radius saturates at a finite value $D_{\rm sat}$, which depends on $R$ and satisfies $D_{\rm sat}<R$.

To determine $D_{\rm sat}$ for a finite system, we integrate the continuity equation, Eq.~\eqref{eq4}, over the 
membrane area. 
Under radial symmetry, mass conservation gives
\begin{equation}
\int_0^R {\rm d r}\, r \phi(r,t)= {\rm const.}
\label{eq9}
\end{equation}
Neglecting the finite interfacial width $\xi$ between the inner and outer domains, 
we obtain the approximate relation
\begin{equation}
\phi_{\rm b}R^2 \approx \phi_{\rm s}D_{\rm sat}^2 -\phi_{\rm s}\left(R^2-D_{\rm sat}^2\right).
\label{eq10}
\end{equation}
Consequently, mass conservation in a membrane of finite radius gives
\begin{equation}
\frac{D_{\rm sat}}{\xi} = \left( \frac{\phi_{\rm s}+\phi_{\rm b}} {2\phi_{\rm s}} \right)^{1/2} \frac{R}{\xi}.
\label{eq11}
\end{equation}
For the parameter values used here, Eq.~\eqref{eq11} gives
$D_{\rm sat}/\xi=38.3<R/\xi=100$, in agreement with the numerical concentration profiles shown in Fig.~\ref{fig3}.

%%%%%%%%%%%%%%%%%%%
\subsection{Domain-growth laws}
\label{sec:growthlaw}

\begin{figure}[tbh]
\centering
\includegraphics[width=0.8\linewidth]{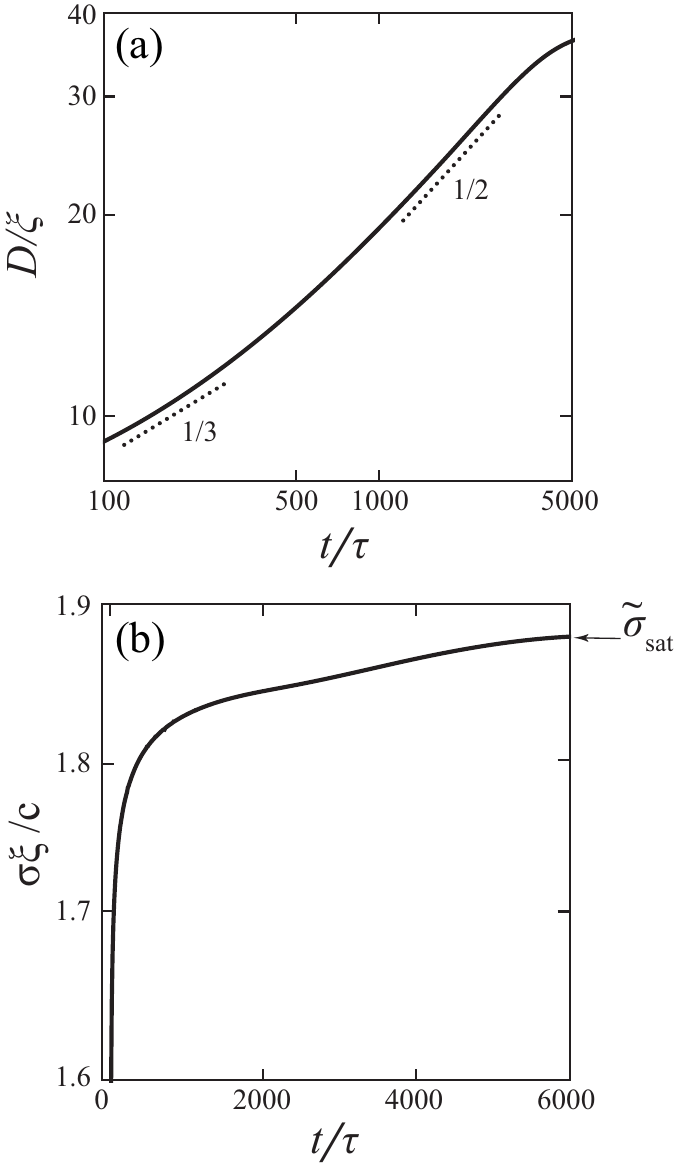} 
\caption{
\textsf{
(a) The dimensionless domain radius $D(t)/\xi$ as a function of the rescaled time $t/\tau$ at intermediate times, 
shown on a log-log scale. 
The parameters are $a/b=\phi_{\rm s}^2=2$ and $R/\xi=100$. 
The dotted lines indicate power-law behavior with exponents $1/3$ and $1/2$ (see the text for details). 
The onset of the plateau associated with $D_{\rm sat}$ becomes apparent at approximately 
$t/\tau=6{,}000$.
(b) The dimensionless line tension $\tilde{\sigma}(t)=\sigma(t)\xi/c$ as a function of the rescaled time 
$t/\tau$ at intermediate times.
At later times, the line tension approaches the saturation value $\tilde{\sigma}_{\rm sat}\approx1.88$, 
in close agreement with the equilibrium planar-interface value $\tilde{\sigma}_0=1.89$.
}}
\label{fig4}
\end{figure}

As shown in Fig.~\ref{fig4}(a), the domain radius exhibits two approximate growth regimes,
$D(t)\sim t^\alpha$, with $\alpha \approx 1/3$ at earlier times and
$\alpha \approx 1/2$ at later times~\cite{2010_Liang,2010_Stanich,2013_Stanich,2012_Komura}.
The crossover can be interpreted in terms of the effective chemical-potential difference that drives
the radial transport of the conserved order parameter.

In Sec.~\ref{sec:circular}, we introduced the local chemical potential as $\mu=\delta F/\delta\phi$.
For a circular domain with a narrow interface, the value of this field in the interfacial region can be approximated 
by a sharp-interface expression. 
Following Ref.~\cite{1994_Bray}, we denote this effective interfacial chemical potential by $\mu_{\rm int}$ and write
\begin{eqnarray}
\mu_{\rm int} = \frac{1}{\Delta\phi} \left(\Delta V-\frac{\sigma}{D}\right),
\label{eq12}
\end{eqnarray}
where $\Delta\phi$ is the jump in the order parameter across the interface, $\Delta V$ is the difference in 
bulk free-energy density between the inner and outer domains, and $\sigma$ is the line tension.

%%%%%%%%%%%%%%%%%%%%%%%%%%%%%%%%%%%%%%
%%%%%%%%%%%%%%%%%%%%%%%%%%%%%%%%%%%%%%%

At early times, the domain radius $D$ is small, and the curvature 
contribution dominates because
$\Delta V\ll\sigma/D$. 
Note that this ``line curvature" relative to the boundary must be distinguished from the membrane curvature. 
Thus, $\lvert\mu_{\rm int}\rvert\sim\sigma/D$. 
Estimating the chemical-potential gradient over a distance of order $D$, we obtain
$\dot{D}  \sim j \sim \nabla \lvert \mu_{\rm int}\rvert \sim \sigma/D^2$,
which yields the growth law 
\begin{equation}
D(t) \sim(\sigma t)^{1/3}.
\end{equation}

At later times, as the domain grows, the bulk contribution becomes dominant and
$\mu\sim\Delta V$ is approximately constant. The corresponding chemical-potential gradient scales as
$\nabla \lvert \mu_{\rm int} \rvert\sim\Delta V/D$, giving
$\dot{D} \sim \Delta V/D$.
The late-time growth law is therefore
\begin{equation}
D (t) \sim(\Delta V t)^{1/2}.
\end{equation}

We conclude that during the early stages of growth, the evolution is governed primarily by the 
increase in the line tension $\sigma$ as the domain curvature decreases; we refer to this as the 
line-tension regime, as explained next. 
At later stages, the increase in the domain radius $D$ becomes the dominant contribution, defining the 
domain-size regime. 

%%%%%%%%%%%%%%%%%%%%%%%%%%
\subsection{Time evolution of the line tension}
\label{sec:line}

The free energy associated with a given concentration profile $\phi(r,t)$ contains contributions from both 
the formation of the curved interface and the bulk free-energy difference $\Delta V$ across it. 
To apply the budding criterion of Ref.~\cite{1992_Lipowsky}, which is formulated in terms of the equilibrium 
line tension, we must define the interfacial contribution carefully. 
In Appendix~\ref{tlt}, we evaluate the line tension $\sigma(t)$ of a growing 2D domain and show 
that it can be given by
\begin{eqnarray}
\sigma(t)=c \int_0^R {\rm d} r \, \left(\frac{\partial\phi(r,t)}{\partial r}\right)^2. 
\label{eq13}
\end{eqnarray}
The time-dependent line tension, $\sigma(t)$, is calculated from Eq.~\eqref{eq13} using the evolving 
concentration profile $\phi(r,t)$. 
Figure~\ref{fig4}(b) shows the corresponding dimensionless line tension,
$\tilde{\sigma}(t)=\sigma(t)\xi/c$, for $0 \leq t/\tau \leq 6{,}000$. 
The line tension approaches a saturation value of approximately $\tilde{\sigma}_{\rm sat}\approx1.88$.

As we see in Fig.~\ref{fig3}, $\phi(r,t)$ varies appreciably in two spatial regions. 
The dominant variation occurs across the interface, around $r\sim D$, while a weaker and more gradual 
variation extends through the outer region, $r>D$. 
Hence, the line tension defined in Eq.~\eqref{eq13} is governed primarily by the interfacial contribution, 
with only a minor contribution from the concentration variation associated with the bulk free-energy difference 
outside the inner domain.

As a consistency check, we first integrate over the full interval $0\leq r/\xi\leq100$ in Eq.~\eqref{eq13}. 
We then restrict the integration range to $0\leq r/\xi\leq r(\phi_{\rm min})/\xi$, where $\phi_{\rm min}$ 
denotes the local minimum located just outside the inflection point. 
The difference in the resulting $\sigma(t)$ using the two procedures is found to be negligible.
This confirms that the dominant contribution to the line tension arises from the concentration variation 
across the A/B interface centered at $r=D$.

In the limit $D\to\infty$, the curved interface becomes locally planar, and our model approaches the 
corresponding 1D equilibrium problem, for which the line tension can be evaluated analytically~\cite{1994_Bray}. 
The equilibrium profile of the 1D Ginzburg-Landau model is 
$\phi(r)=-\phi_{\rm s}\tanh [r/(\sqrt{2}\xi)]$,
which satisfies
$\phi(-\infty)=\phi_{\rm s}$ and
$\phi(\infty)=-\phi_{\rm s}$.
Substituting this profile into the expression analogous to Eq.~\eqref{eq13}, we obtain the dimensionless 
planar line tension $\tilde{\sigma}_0=1.89$. 
This value agrees closely with the numerically obtained saturation value, $\tilde{\sigma}_{\rm sat}=1.88$.
In Appendix~\ref{sec:gro}, we examine the radius dependence of the line tension 
and estimate the resulting Tolman length~\cite{1948_Tolman, 1949_Tolman}.

%%%%%%%%%%%%%%%%%%%%%%%%%%%%%%%%%%
\begin{figure}[tbh]
\centering
\includegraphics[width=0.7\linewidth]{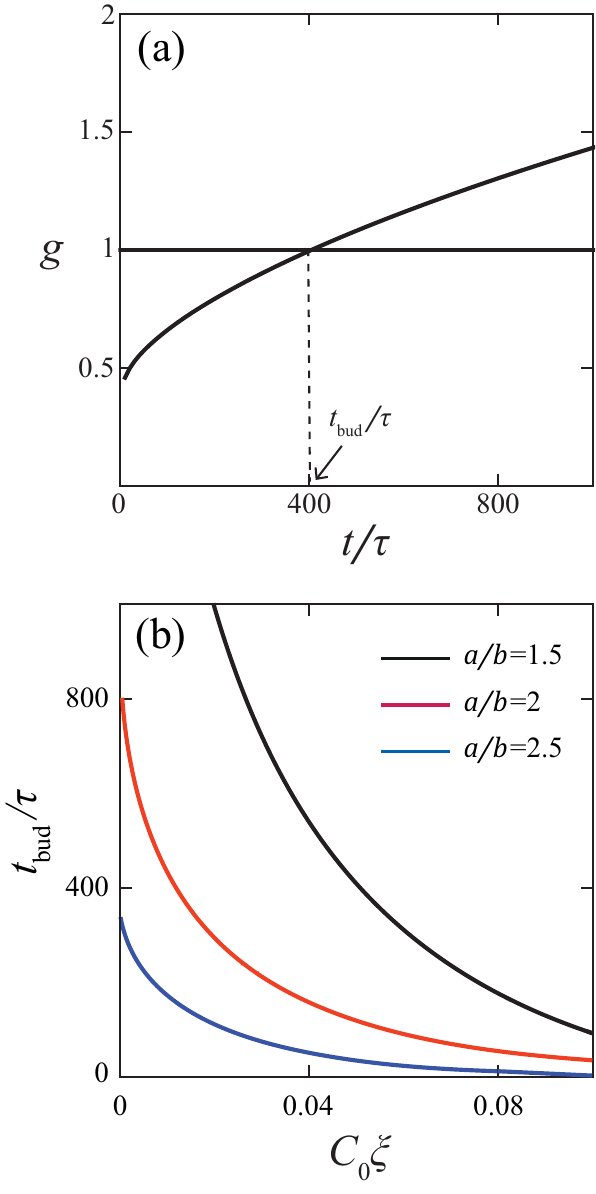} 
\caption{
\textsf{
(a) The dimensionless budding function $g(t)$ defined in Eq.~\eqref{eq16}, together with the threshold value $g=1$. 
Their intersection determines the budding-instability time $t_{\rm bud}/\tau$. 
The parameters are $a/b=\phi_{\rm s}^2=2$, $C_0\xi=0.01$, and $\kappa/c=4$, and 
we obtain $t_{\rm bud}/\tau \approx 400$.
The estimated budding-instability time lies within the growth regime characterized by $D\sim t^{1/3}$.
(b) The rescaled buddingtime $t_{\rm bud}/\tau$ as a function of the dimensionless spontaneous curvature 
$C_0\xi$ for $a/b=\phi_{\rm s}^2=1.5$, $2$, and $2.5$, over the range $0\leq C_0\xi\leq0.1$.
}}
\label{fig5}
\end{figure}
%%%%%%%%%%%%%%%%%%%%%%%%%%

%%%%%%%%%%%%%%%%
\section{Budding instability}
%%%%%%%%%%%%%%%%
\label{sec:lipo}

Next, following Ref.~\cite{1992_Lipowsky}, we consider domain-induced budding. 
The total energy of a partially budded domain with out-of-plane curvature $C$, 
connected to an otherwise planar membrane by a neck with line tension $\sigma$, is given by
\begin{equation}
E = 2\pi\sigma D \sqrt{1-\left(\frac{DC}{2}\right)^2} + 2\pi\kappa \left(DC-DC_0\right)^2.
\label{eq14}
\end{equation}
The first term represents the line-energy contribution associated with the neck. In contrast, 
the second term represents the bending energy of the bud and depends on the bending 
rigidity $\kappa$ and the membrane spontaneous curvature $C_0$.
The critical radius is determined by the loss of local stability of the partially budded state. 
Accordingly, the conditions $\partial E/\partial C=\partial^2 E/\partial C^2=0$ are imposed simultaneously, yielding
\begin{equation}
D^*=\frac{8\kappa/\sigma} {\left[ 1+\left(4C_0\kappa/\sigma\right)^{2/3} \right]^{3/2}}.
\label{eq15}
\end{equation}
For $C_0\neq 0$, the planar state $C=0$ is not an extremum of the
energy, and the domain is therefore generally weakly deformed before
the instability is reached.
Equation~(\ref{eq15}) should thus be interpreted as the loss of local stability
of the partially budded state toward a more strongly budded configuration.

A lipid domain may possess a nonzero spontaneous curvature $C_0$ when the up-down symmetry 
of the bilayer is broken. 
In particular, compositional asymmetry between the two leaflets can generate an effective spontaneous 
curvature of the bilayer domain through curvature-composition coupling. 
In a perfectly symmetric bilayer, by contrast, the spontaneous curvatures of the two monolayers cancel, 
resulting in a vanishing net spontaneous curvature of the bilayer.
The present Model B description contains only a single in-plane composition field and does not resolve the 
two leaflets separately. 
Therefore, the spontaneous curvature $C_0$ is not generated by the domain-growth model itself, 
but is introduced phenomenologically to represent an underlying leaflet asymmetry or another microscopic curvature-generating 
mechanism.

The connection between our 2D domain-growth dynamics and the budding criterion described above is 
established through a quasistatic approximation. 
We assume that the domain grows slowly through a sequence of states, each characterized by an instantaneous 
radius $D(t)$ and line tension $\sigma(t)$. 
We further assume that the time evolution of $D(t)$ and $\sigma(t)$ is unaffected by the spontaneous curvature $C_0$.
We define the budding-instability time $t_{\rm bud}$ as the earliest time at which the evolving domain reaches the critical 
condition in Eq.~(\ref{eq15}).
Because line tension varies during domain growth, we must evaluate the critical condition using its instantaneous value. 
At $t=t_{\rm bud}$, we therefore require
\begin{equation}
\frac{D(t_{\rm bud})}{8\kappa/\sigma(t_{\rm bud})} 
\left[ 1+ \left( \frac{4C_0\kappa}{\sigma(t_{\rm bud})} \right)^{2/3} \right]^{3/2}=1.
\label{eq15a}
\end{equation}

To determine $t_{\rm bud}$ numerically, we define the dimensionless time-dependent function
\begin{equation}
g (t) = \frac{D(t)}{8\kappa/\sigma(t)} \left[ 1+ \left(\frac{4C_0\kappa}{\sigma(t)} 
\right)^{2/3} \right]^{3/2}, 
\label{eq16}
\end{equation}
which is plotted in Fig.~\ref{fig5}(a).
The budding-instability time $t_{\rm bud}$ is defined as the earliest time at which $g(t_{\rm bud})=1$.
For the parameter values
$a/b=2$,
$C_0 \xi=0.01$, and 
$\kappa/c=4$, we obtain
$t_{\rm bud}/\tau \approx 400$.
This time lies within the growth regime characterized by
$D\sim t^{1/3}$.
At the estimated budding threshold, the corresponding dimensionless radius and line tension are
$D(t_{\rm bud})/\xi \approx 13.6$ 
and 
$\sigma(t_{\rm bud})\xi/c \approx 1.80$.

Figure~\ref{fig5}(b) shows the rescaled budding-instability time, $t_{\rm bud}/\tau$, as a function of the dimensionless 
spontaneous curvature $C_0\xi$ for several values of $a/b=1.5$, $2$, and $2.5$. 
The budding-instability time is calculated from Eq.~\eqref{eq15a} for domains with small spontaneous 
curvature, $0<C_0\xi\leq0.1$. 
For each fixed value of $a/b$, the budding-instability time decreases as $C_0\xi$ increases, because a larger 
spontaneous curvature lowers the critical domain size required for the budding instability.
Moreover, at fixed $C_0\xi$, $t_{\rm bud}$ increases as the system approaches the critical point, corresponding 
to smaller $a/b$.

%%%%%%%%%%%%%%%%%%%%
\section{Summary and discussion}
%%%%%%%%%%%%%%%%%%%%
\label{sec:dis}

In this paper, we have investigated how the nonequilibrium growth of a circular domain in a phase-separating 
lipid membrane brings the system toward a budding instability.
Using a conserved time-dependent Ginzburg-Landau model under radial symmetry, we calculated the evolving concentration profile. 
We extracted the domain radius $D(t)$ and the time-dependent line tension $\sigma(t)$ from it.
The domain radius exhibits approximate growth laws, $D(t)\sim t^{1/3}$ at earlier times and $D(t)\sim t^{1/2}$ at later times, 
before reaching a finite-size saturation set by conservation of the total order parameter.
In contrast, the line tension rises rapidly during the early evolution and subsequently approaches the equilibrium planar-interface value.
The different time dependences of $D(t)$ and $\sigma(t)$ lead to a crossover from an early line-tension regime, 
in which the evolving interface plays the dominant role, to a later domain-size regime, in which the continued growth 
of the domain becomes increasingly important.

To connect this planar growth dynamics to membrane budding, we combined the calculated $D(t)$ and $\sigma(t)$ 
with Lipowsky's equilibrium instability criterion within a quasistatic approximation.
The spontaneous curvature $C_0$, introduced phenomenologically to represent leaflet asymmetry, enters the budding criterion 
but is assumed not to affect the preceding in-plane growth dynamics.
The budding-instability time is then determined as the earliest time at which the instantaneous values of $D(t)$ and $\sigma(t)$  satisfy the instability condition.
In this description, instability occurs earlier for larger spontaneous curvature, whereas it is delayed as the system approaches the critical point.
Overall, our results show that the coupled evolution of domain size and interfacial properties governs the budding instability.

The present model does not describe the membrane-shape dynamics after the instability and neglects the 
feedback of curvature-composition coupling on the pre-budding evolution. 
A more complete description would couple the conserved composition field directly to membrane shape, 
for example, through a composition-dependent spontaneous curvature. 
Such an extension would make it possible to determine the actual budding dynamics rather than only the 
estimated onset of the instability.

\smallskip
%%%%%%%%%%%%%%%%
\noindent {\bf Acknowledgements}~~~
This manuscript honors the memory of Wolfgang Helfrich, whose groundbreaking concepts 
fundamentally shaped the physics of soft matter and biomembranes. 
The mathematical treatments of membrane bending elasticity and spontaneous curvature
used in this paper are heavily indebted to his seminal theories. 
His profound contributions to membrane biophysics continue to inspire our field.
Additionally, Helfrich co-invented (with M. Schadt) the Twisted Nematic (TN) 
field effect in 1970, which later found its way into billions of LCD devices worldwide.

J.W.\ acknowledges support from the French ORT association and the ORT School
of Strasbourg. 
D.A.\ acknowledges partial support from the Israel Science Foundation (ISF) (Grant No.\ 226/24). 
H.D.\ acknowledges support from the ISF (Grant No. 1611/24).
S.K.\ acknowledges support from the National Natural Science Foundation of China (Grant No.\ 12274098) and 
from the Zhejiang Key Laboratory of Soft Matter Biomedical Materials (2025ZY01036 and 2025E10072).
%\end{acknowledgements}

\medskip
%%%%%%%%%%%%
\noindent{\bf Declaration of Interests}\\
The authors declare no competing interests.

\appendix
%%%%%%%%%%%%%%%%%%%%%%%%%%%%%%%%%%
\section{Role of hydrodynamics in isotropic domain growth}
%%%%%%%%%%%%%%%%%%%%%%%%%%%%%%%%%%
\label{he} 

The thermodynamic driving force per unit area is
${\mathbf g}(\mathbf r) = -\nabla (\delta F/\delta\phi)$.
It is localized near the domain boundary, where the concentration gradient is significant. 
For the circular geometry considered here, the force is directed normal to the boundary and can be written as
${\mathbf g}(\mathbf r) = g_0(\mathbf r) \hat{\mathbf n}(\mathbf r)$,
where $\hat{\mathbf n}$ is the unit normal vector. 
Through membrane hydrodynamics, the forces acting at points $\mathbf r'$ along the boundary generate a flow 
velocity at another position $\mathbf r$ according to
\begin{align}
{\mathbf v}(\mathbf r)
& = \int {\rm d}{\mathbf r'} \, \mathbf{G}(\mathbf r-\mathbf r') \cdot{\mathbf g}(\mathbf r')
\nonumber \\
& = \int {\rm d}{\mathbf r'} \, g_0(\mathbf r') \mathbf{G}(\mathbf r-\mathbf r') \cdot
\hat{\mathbf n}(\mathbf r'),
\end{align}
where $\mathbf{G}$ is the hydrodynamic Green's function for the membrane, analogous to the Oseen 
tensor in a 3D fluid.

Since the problem is isotropic, the force magnitude is uniform along the circular boundary, so that
\begin{equation}
{\mathbf v}(\mathbf r) = g_0 \int {\rm d} {\mathbf r'} \, \mathbf{G}(\mathbf r-\mathbf r') 
\cdot\hat{\mathbf n}(\mathbf r').
\end{equation}
Using the divergence theorem, this expression can be rewritten as
\begin{equation}
{\mathbf v}(\mathbf r) = g_0 \int_{R > r'} {\rm d} {\bf R} \, \nabla_{\mathbf R} \cdot 
\mathbf{G}(\mathbf r- \mathbf R).
\end{equation}
For an incompressible membrane flow, the hydrodynamic Green's function is divergence-free,
$\nabla_{\bf R}\cdot\mathbf{G}({\bf r}-{\bf R})=0$.
Therefore, the flow contributions from different points on the circular interface cancel, resulting in zero 
net advection.

%%%%%%%%%%%%%%%%%%%%%%%%%%%%
\section{Definition of the dynamical line tension}
%%%%%%%%%%%%%%%%%%%%%%%%%%%%
\label{tlt}

Assuming, as throughout this work, that the free-energy functional in Eq.~\eqref{eq1} remains applicable 
out of equilibrium, we define the line tension in the usual way as the free-energy cost per unit length 
required to create an interface between the two domains.

At equilibrium, the free-energy densities of the two coexisting phases are equal, so that
$V(\phi_{\rm s})=V(-\phi_{\rm s})$. 
For two semi-infinite equilibrium domains separated by a straight interface at $x=0$, the excess 
free energy per unit length associated with the interface is
\begin{equation}
\sigma_{\rm eq}=\int_{-\infty}^{\infty} \mathrm{d}x 
\left[ \frac{c}{2} \left(\frac{\mathrm{d}\phi}{\mathrm{d}x}\right)^2 + V(\phi)-V(\pm\phi_{\rm s}) \right].
\end{equation}
Minimizing the free energy with respect to the profile $\phi(x)$ and substituting the resulting equilibrium 
profile back into the integral yields~\cite{1994_Bray}
\begin{equation}
\sigma_{\rm eq} = c\int_{-\infty}^{\infty} \mathrm{d}x
\left(\frac{\mathrm{d}\phi}{\mathrm{d}x}\right)^2.
\label{eq21}
\end{equation}

%%%%%%%%%%%%%%%%%%%%%%%%%%%%%%%%%%%%%%%%%%%%%
\begin{figure}[tbh]
\centering
\includegraphics[width=0.8\linewidth]{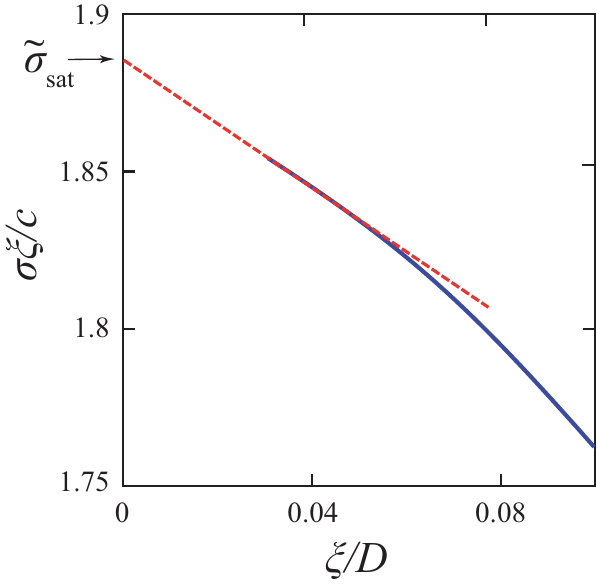}   
\caption{\textsf{
The dimensionless line tension $\tilde{\sigma}=\sigma\xi/c$ as a function of the dimensionless 
inverse domain radius $\xi/D$. 
The solid blue line shows the numerical results, while the red dashed line represents a fit to the asymptotic 
Tolman expression in Eq.~\eqref{eq_Tol}. 
The fit yields the planar-interface value $\tilde{\sigma}_{\rm sat}\approx1.88$ in the limit 
$\xi/D\to 0$ and a Tolman length $\delta/\xi\approx0.5$.
}}
\label{fig6}
\end{figure}
%%%%%%%%%%%%%%%%%%%%%%%%%%%%%%%%%%%%%%%%%%%%%%%%

Our system differs from this simple case in two aspects: (i) the domains are finite, and (ii) the system 
is out of equilibrium, with $V(\phi_{\rm s})<V(\phi_{\rm b})$. 
We propose that, in this case, the line tension is well approximated by
\begin{equation}
\sigma = c\int_{0}^{R} \mathrm{d}r \left(\frac{\partial\phi}{\partial r}\right)^2,
\label{sigmad}
\end{equation} 
which we adopt in the main text [see Eq.~\eqref{eq13}].

To provide a rigorous basis for Eq.~\eqref{sigmad}, let us consider an interface of width $\xi\ll D$. 
We assume, approximately, that $\phi(r)=\phi_{\rm s}$ for $r<D-\xi$ and $\phi(r)=\phi_{\rm b}$ for $r>D+\xi$. 
The free energy can then be written as
\begin{align}
    F & = \pi(D-\xi)^2 V(\phi_{\rm s}) +
    \pi[R^2-(D+\xi)^2] V(\phi_{\rm b})
    \nonumber\\
    & \quad + \ 2\pi D \int_{D-\xi}^{D+\xi} {\rm d}r \left[ \frac{c}{2} \left( \frac{\partial\phi}{\partial r} \right)^2 + V(\phi) \right] 
    \nonumber\\
    & = F_{\rm bulk} + 2\pi D \sigma,
\end{align}
where $F_{\rm bulk} = \pi D^2 V(\phi_{\rm s}) + \pi(R^2-D^2) V(\phi_{\rm b})$, and
\begin{equation}
    \sigma \approx \int_{D-\xi}^{D+\xi} {\rm d}r \left[ \frac{c}{2} \left( \frac{\partial\phi}{\partial r} \right)^2 + V(\phi) \right] 
    - \xi[ V(\phi_{\rm s})+V(\phi_{\rm b})].
\label{sigmad2}
\end{equation}

Within the narrow interfacial region, we assume that the concentration profile is locally close to an equilibrium profile. 
Under this approximation, the profile satisfies
$c (\partial^2\phi/\partial r^2) \approx V'(\phi)$,
and a first integration gives
$(c/2) \left(\partial\phi/\partial r\right)^2 \approx V(\phi)+{\rm const.}$
By requiring the gradient to vanish on the inner side of the interface, $r<D-\xi$ (see Fig.~\ref{fig3}), where $\phi \approx \phi_{\rm s}$, 
the integration constant is approximately $-V(\phi_{\rm s})$. 
Hence,
$ V(\phi) \approx (c/2) \left(\partial\phi/\partial r\right)^2 +V(\phi_{\rm s})$.
Substituting this relation into Eq.~\eqref{sigmad2}, we obtain
\begin{equation}
\sigma \approx 
c\int_{0}^{R}{\rm d}r \left(\frac{\partial\phi}{\partial r}\right)^2
-\xi\left[V(\phi_{\rm b})-V(\phi_{\rm s})\right].
\end{equation}
Here, the integration range has been extended approximately to the entire system, 
since the concentration gradient outside the interface is much smaller than that within the interface.
Thus, Eq.~\eqref{sigmad} is recovered up to a small negative correction that vanishes in the equilibrium limit,
$V(\phi_{\rm b})\to V(-\phi_{\rm s})=V(\phi_{\rm s})$.

%%%%%%%%%%%%%%%%%%%%%%%%
\section{Line tension and Tolman length}
%%%%%%%%%%%%%%%%%%%%%%%%
\label{sec:gro}

In earlier work, Tolman~\cite{1948_Tolman,1949_Tolman} showed that the surface tension of a finite-sized domain 
in three dimensions (3D) depends on the curvature of its interface and generally increases as the curvature decreases. 
By analogy, we apply Tolman’s result to 2D systems by replacing the 3D surface tension with a 2D line tension. 

As discussed in Sec.~\ref{sec:line}, the line tension represents the energetic cost associated with the 
domain boundary. 
Following Refs.~\cite{1948_Tolman,1949_Tolman,1993_Blokhuis,2006_Blokhuis,2019_Dimova}, the curvature 
dependence of the line tension can be expressed as an expansion in the dimensionless ratio $\xi/D$, 
where $\xi$ is the interfacial width and $D$ is the domain radius:
\begin{equation}
\sigma = \sigma_0 \left[ 1-c_1\frac{\xi}{D} -c_2\left(\frac{\xi}{D}\right)^2 +\cdots \right].
\end{equation}
Here, $\sigma_0$ is the planar-interface value obtained in the limit $D\to\infty$, while $c_1$ and $c_2$ 
are dimensionless coefficients characterizing the spontaneous curvature and the bending rigidity corrections, respectively.
Introducing the Tolman length $\delta$ through $\delta= c_1\xi/2$ related to the spontaneous curvature and $\delta'= c_2\xi^2$ 
related to the bending rigidity, the expansion to the second order becomes
\begin{equation}
\sigma \approx \sigma_0 \left( 1-\frac{2\delta}{D}-\frac{\delta'}{D^2} \right).
\label{eq_Tol} 
\end{equation}

Figures~\ref{fig4} and \ref{fig5} show the time dependence of $D(t)$ and $\sigma(t)$, respectively. 
To examine the consistency with Tolman's curvature correction, we eliminate time and express the line 
tension as a function of the domain radius, $\sigma=\sigma(D)$. 
Figure~\ref{fig6} shows the dimensionless line tension, $\tilde{\sigma}=\sigma \xi/c$, as a function of the 
dimensionless inverse radius $\xi/D$.
By fitting the asymptotic form in Eq.~\eqref{eq_Tol} in the small-curvature regime, $\xi/D\ll1$, we estimate 
the dimensionless Tolman length to be $\delta/\xi\approx0.5$, in reasonable agreement with previous 
studies~\cite{2012_Kolomietz} and the bending rigidity $\delta'/\xi^2\approx1.4$. 
The extracted Tolman length is of the same order as the interfacial width $\xi$, consistent with its interpretation 
as a microscopic length scale determined by the interfacial structure. It is worth noting that the bending rigidity 
governs the fluctuations of the domain shape, which are even higher as the bending rigidity decreases.

%%%%%%%%%%%%%%%%

\end{document}